\documentclass[pra,letterpaper,onecolumn,showpacs,superscriptaddress,floatfix]{revtex4}
\usepackage{graphicx,psfrag,amsmath,amssymb,amsfonts,bbm,latexsym,color,dcolumn,bm}

\begin{document}
\title{Modelling superradiant amplification of Casimir photons \\ 
in very low dissipation cavities}

\author{J.H. Brownell}

\affiliation{Department of Physics and Astronomy, Dartmouth College, 6127 Wilder Laboratory,
Hanover, NH 03755, USA}

\author{W.J. Kim}

\affiliation{Department of Physics and Astronomy, Dartmouth College, 6127 Wilder Laboratory,
Hanover, NH 03755, USA}

\author{R. Onofrio}

\affiliation{Department of Physics and Astronomy, Dartmouth College, 6127 Wilder Laboratory,
Hanover, NH 03755, USA}

\affiliation{Dipartimento di Fisica ``Galileo Galilei'', Universit\`a di Padova,
Via Marzolo 8, Padova 35131, Italy}

\begin{abstract}
Recent advances in nanotechnology and atomic physics may 
allow for a demonstration of the dynamical Casimir effect. 
An array of film bulk acoustic resonators (FBARs) coherently driven at 
twice the resonant frequency of a high quality electromagnetic 
cavity can generate a stationary state of Casimir photons. 
These are detected using an alkali atom beam prepared in an 
inverted population of hyperfine states, with an induced superradiant 
burst producing a detectable radio-frequency signal. 
We describe here the results of the simulations of the 
dynamics of superradiance and superfluorescence, with the aim 
to optimize the parameters for the detectability of Casimir photons.
When the superradiant lifetime is shorter than the dissipation time,
we find superradiant evolution to be similar in character but
dramatically slower than in the usual lossy case. 
\end{abstract}

\pacs{12.20.Fv, 42.50.Pq, 85.85.+j, 42.50.Lc}

\maketitle

\section{Introduction}
Observable effects due to the change in the boundary conditions of quantum fields, like
the creation of particles in an expanding universe \cite{Schroedinger} or the Casimir force
\cite{Casimir}, provide crucial information on quantum vacuum at the macroscopic level.
After the recent results on Casimir forces, with measurements performed in a variety of
geometries ranging from the original parallel plane \cite{Sparnaay,Bressi} to the
sphere-plane \cite{vanBlokland,Lamoreaux,Mohideen,Iannuzzi,Decca} and crossed-cylinders
\cite{Ederth}, there is interest to understand dissipative effects of vacuum fluctuations,
especially its interplay with relativity \cite{Moore,Fulling,Jaekel,Lambrecht}.
This dissipation mechanism should induce irradiation of photons, a phenomenon also
known as dynamical Casimir effect \cite{Barton,Dodonov,Plunien,Crocce}.
This can be understood both as the creation of particles under non-adiabatic 
changes in the boundary conditions of quantum fields, or as classical parametric 
amplification with the zero point energy of a vacuum field mode as input state. 
In this paper, following on the proposal described in \cite{Andy}, we describe 
a model for the superradiant amplification scheme with particular emphasis on 
its dynamics and the optimization of the involved parameters.

\section{Generation and detection of Casimir photons}

As discussed in more detail in \cite{Andy}, under parametric amplification in 
an electromagnetic cavity an initial state of $N_0$ photons with frequency 
within the resonance bandwidth of the fundamental mode of the cavity 
$\omega$ is transformed into a squeezed state with an average number 
of photons growing in time as \cite{Dodonov,Plunien,Crocce}:
\begin{equation} 
N_{\mathrm{Cas}}(t)=N_0 \sinh^{2}(\omega_{\mathrm{mech}} \epsilon t),
\label{eq:photonnumber}
\end{equation}
provided that the parametric resonance condition with a mechanical driving at a
frequency $\omega_{\mathrm{mech}}=2 \omega$ is fulfilled. The term $\omega_{\mathrm{mech}} 
\epsilon$ in the hyperbolic sine function represents the squeezing parameter, with the modulation 
depth $\epsilon=v/c$, where $v$ is the velocity of the resonator and $c$ the speed of light. 
This exponential growth is eventually limited by the photon leakage of the cavity 
expressed through its quality factor $Q$, which saturates
at the hold time $\tau=Q/\omega$, reaching a maximum photon population:
\begin{equation}
N_{\mathrm{Cas}}^{\mathrm{max}}=N_{\mathrm{Cas}}(\tau)=N_0 \sinh^{2}(2 Q \epsilon).
\label{eq:sat}
\end{equation}
Given an initial number of photons in the cavity, which will be attributable 
to the quantum vacuum at temperatures such that $K_B T  \ll \hbar \omega$ 
(see also \cite{EPLcomment} for a detailed discussion), the average number 
of photons at saturation in Eq.~(\ref{eq:sat}) strongly depends on the 
product of two parameters, $Q$ and $\epsilon$, which can be on the 
order of $10^8$ and $10^{-8}$ respectively. 
The expected saturated power initiated by Casimir emission is:
\begin{equation} 
P_{\mathrm{Cas}}=
N_{\mathrm{Cas}}^{\mathrm{max}} \frac{\hbar \omega}{\tau}
\label{eq:phpower}
\end{equation}
and for a 3.0 GHz FBAR resonator and $Q\epsilon \simeq 1$, the saturated
power is $3 \times 10^{-22}$ W, which is too low to be detectable using current technology.
This demands the use of an efficient, nearly quantum-limited, photon detector in the
radio-frequency range. Ultra-sensitive atomic detection schemes can be exploited for
detecting Casimir photons by preparing an ensemble of population-inverted atoms in a
particular hyperfine state, which for alkali atoms ranges from 0.2 GHz for Li to 9 GHz for Cs,
whose transition frequency corresponds to the cavity resonance. An additional amplification
process is available in which the weak Casimir signal triggers the stimulated emission of
the ensemble of atoms. This effect is a form of superradiance \cite{Benedict,Gross}.
The hyperfine transition in the ground state occurs through a magnetic dipole interaction,
and its natural lifetime in free space is approximately:
\begin{equation} 
T_1 \approx \frac{4 \pi}{\mu_0} \frac{3 \hbar}{4 \mu_{B}^2 (\omega/c)^3},
\label{eq:T1}
\end{equation}
where $\mu_{B}$ is the Bohr magneton and $\mu_0$ the magnetic permeability in vacuum.
This natural lifetime in free space is favorably reduced inside a resonant cavity due to
the modification of density of states \cite{Purcell,Goy}:
\begin{equation}
T_1^\mathrm{cav} = \frac{4\pi^2}{3Q}
\frac{V}{\lambda^3} \, T_1 \approx \frac{4 \pi}{\mu_0}
\frac{\hbar \, V}{8\pi \mu_{B}^2 Q},
\label{eq:Tcav}
\end{equation}
where $V$ is the cavity mode volume. For a few GHz cavity with 1~cm$^2$ cross-sectional 
area and $Q=10^8$, the natural lifetime is reduced by a factor of $10^{10}$. 
In spite of this cavity-enhanced spontaneous rate, the typical hyperfine transition 
lifetime for the alkali atoms is still impractically long, on the order of $10^3 \div 10^{5}$~s.
To shorten this timescale, let us suppose to inject $N_{\mathrm{at}}$ excited atoms into the cavity.  
The Casimir field acts on all atoms, stimulating emission on a time scale of the superradiant lifetime, 
defined as $T_{\mathrm{SR}} = T_1^{\mathrm{cav}}/N_{\mathrm{at}}$, which is in the millisecond range 
for $N_{\mathrm {at}} \approx 10^8$ or less. An atomic density large enough will then induce 
a superradiant burst with peak power of $P_{\mathrm {SR}} \approx N_{\mathrm{at}} {\hbar \omega}/{T_{\mathrm {SR}}}$,
increasing quadratically with the number of atoms. Considering a few GHz resonator with $10^8$ atoms 
and $T_{\mathrm {SR}}=10^{-3}$ s, yields $P_{\mathrm{SR}} = 10^{-13}$~W, a billionfold improvement 
over the power without superradiant amplification as in Eq.~(\ref{eq:phpower}).

Spontaneous emission into the cavity mode by the atoms will also trigger a superradiant burst,
a process also known as superfluorescence. To distinguish this source of background from the Casimir stimulated
superradiance signal, one may study the temporal intensity envelope of the amplified photons. 
Both the average delay ($T_{\mathrm {D}}$) of the peak intensity from the initial excitation of the atomic
population and its fluctuation ($\Delta T_{\mathrm {D}}$) are reduced with increasing number of
resonant photons $N_{\mathrm{ph}}$ initially present \cite{Haroche}:
\begin{equation}
T_{\mathrm {D}}=T_{\mathrm {SR}} \ {\ln} \biggl(
\frac{N_{\mathrm {at}}}{1+N_{\mathrm {ph}}}\biggr),\;\;\;\;\;
\Delta T_{\mathrm {D}}=2 T_{\mathrm {SR}}/\sqrt{1+N_{\mathrm {ph}}}.
\label{eq:delay}
\end {equation}
It should be noted that these standard results follow when the cavity lifetime is much shorter 
than the superradiant lifetime.  For very high $Q$, the system of coupled equations of motion 
must be integrated directly. Tailoring the atomic number can further distinguish the Casimir
stimulated superradiance from superfluorescent pulses. In order for the superradiant pulse 
to develop fully, the growth rate must exceed any decay process, which is primarily due to Doppler
dephasing in the atomic cloud, and the atoms must remain in the interaction region for a time 
longer than the delay time. A proper choice of $N_{\mathrm {at}}$ may suppress superfluorescence 
relative to Casimir superradiance provided that the atoms will be removed from the cavity after 
the expected Casimir delay time but prior to the superfluorescence delay ($N_{\mathrm{ph}}=0$ 
in Eq.~(\ref{eq:delay})).
\begin{figure}[t]
\begin{center}
\includegraphics[width=0.6\columnwidth,clip]{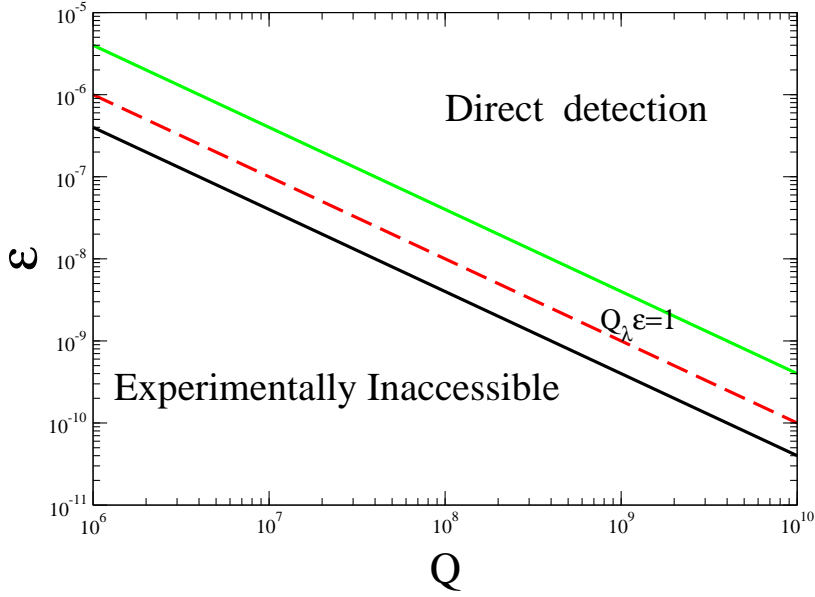}
\caption{Detectability plot in the $\epsilon- Q$ plane.
Depicted from top to bottom are the regions in which direct detection is possible with
state-of-art radiofrequency detectors, the region where
superradiant amplification extends the range of detection, and the
region experimentally inacessible also to superradiant amplification, with
its lower limit due to the speed of micro-bolometers and heterodyne
receivers of the current generation. The dashed line corresponds to the
benchmark values of $Q$ and $\epsilon$ satisfying $Q \epsilon=1$ which 
are at the edge of current technology of superconducting cavities and nanotechnology.}
\end{center}
\label{fig1}
\end{figure}
The superradiant emission can be detected by coupling a power or field detector to the cavity, 
with response time small enough to resolve one superradiant lifetime. 
One issue with this direct measurement is the possible reduction of the quality factor 
of the cavity especially for large coupling efficiency. Micro-bolometers mounted on 
etched ``spider-webs'' have an ultimate sensitivity of 
$10^{-16}\,\mathrm{W}/\sqrt{{\mathrm{Hz}}}$ in the GHz range \cite{Turner}.
Spectrum analyzers are sensitive to sub-fW RF power of kHz bandwidth \cite{Agilent},
and the temporal profile of the burst can be reconstructed through vector analysis.

\section{Superradiant amplification model}

To validate these qualitative estimates, it is necessary to simulate the dynamics of an atomic beam 
travelling through a low dissipation cavity. We shall discuss here specifically Na atoms initially 
optically pumped into the $|F=2, m_F=2>$ ground state, though the approach is general.
Let us consider the second-quantized Hamiltonian with relativistic and hyperfine corrections 
absorbed into the unperturbed atom term $H^{\mathrm{atom}}$ \cite{Eberly}:
\begin{equation}
H = \sum_{j=1}^N  H_j^{\mathrm{atom}} - \frac{e}{m_j} \mathbf{p}_j
\cdot \mathbf{A}(\mathbf{r}_j) + \frac{e^2}{2 m_j} \left|
\mathbf{A}(\mathbf{r}_j)\right|^2 - {\mathbf{\mu}}_j \cdot
\mathbf{B}(\mathbf{r}_j)+ \sum_{\lambda} \hbar \omega_{\lambda} \left(
\hat{a}_{\lambda}^{\dag}\hat{a}_{\lambda} + 1/2 \right).
\label{Hamiltonian}
\end{equation}
The subscripts $j$ and $\lambda$ are the atom and field mode index, respectively, 
$\omega_{\lambda}$ is the mode angular frequency. The fields $\mathbf{A}(\mathbf{r}_j)$ 
and $\mathbf{B}(\mathbf{r}_j)=\nabla \times \mathbf{A}(\mathbf{r}_j)$ are defined 
in terms of field profiles functions $\mathbf{U}(\mathbf{r}_j)$ and creation
and annihilation operators $(a, a^{\dag})$ as:
\begin{eqnarray}
\mathbf{A}(\mathbf{r}_j) &=& \sum_{\lambda} \sqrt{\frac{\hbar}{2 \epsilon_0 \omega_{\lambda}}}
\left[ \mathbf{U}_{\lambda}(\mathbf{r}_j) \hat{a}_{\lambda} +
\mathbf{U}_{\lambda}^{*}(\mathbf{r}_j) \hat{a}_{\lambda}^{\dag} \right]
\\
\mathbf{B}(\mathbf{r}_j) &=& i \sum_{\lambda} \sqrt{\mu_0 \hbar \omega_{\lambda}/2}
\left[ \hat{\mathbf{k}}(\mathbf{r}_j) \times \mathbf{U}(\mathbf{r}_j)
\hat{a}_{\lambda} - \hat{\mathbf{k}}(\mathbf{r}_j) \times \mathbf{U}(\mathbf{r}_j)^{*} \hat{a}_{\lambda}^{\dag} \right]
\end{eqnarray}
\noindent
where the field profile functions form an orthonormal set and the
$a, a^{\dag}$ operators fulfil the usual commutation relationships:
\begin{equation}
\int d^3r \mathbf{U}_{\lambda}(\mathbf{r}_j)\mathbf{U}_{\lambda^{\prime}}^{*}(\mathbf{r}_j) =
\delta_{\lambda\lambda^{\prime}};
\left[ \hat{a}_{\lambda}, \hat{a}_{\lambda^{\prime}}^{\dag} \right] = \delta_{\lambda\lambda^{\prime}};
\left[ \hat{a}_{\lambda}, \hat{a}_{\lambda^{\prime}} \right] =
\left[ \hat{a}_{\lambda}^{\dag}, \hat{a}_{\lambda^{\prime}}^{\dag} \right] = 0. \\
\end{equation}
Given no initial population on an upper state of an electric
dipole transition and no initial or applied field resonant with
the same, the second term can be ignored and each atom evolves
only within the manifold of ground hyperfine states. Also, the
third term is negligible compared to the fourth in this case.
The atomic Hamiltonian then can be represented by an $8 \times 8$
matrix and the Heisenberg equations of motion derived with some
effort. To simplify the discussion, let us assume a cylindrically
symmetric cavity with field propagation ($\hat{\mathbf{k}}$)
primarily along the cavity axis ($\hat{\mathbf{z}}$), with
quantization axis along the cavity axis, so that the active modes 
will be circularly polarized. If we also assume the ideal case 
where the Na atoms are prepared in the $|F=2,m_F=2 \rangle$ state, 
then only the $|F=2, m_F=2 \rangle - |F=1, m_F=1 \rangle$ transition will 
be active and the Hamiltonian can be reduced to:
\begin{equation}
\label{Hamiltonian1}
    H = \sum_{j=1}^N  \left(
\begin{array}{cc}
\hbar \left(\Omega - \mathbf{k} \cdot \mathbf{v}_j \right) & \mu_- B_+(\mathbf{r}_j) \\
\mu_+  B_-(\mathbf{r}_j) & 0 \\
\end{array}
\right) + \sum_{\lambda} \hbar \omega_{\lambda} \left(\hat{a}_{\lambda}^{\dag} \hat{a}_{\lambda} + 1/2 \right),
\end{equation}
using the standard notation for polarization,
\begin{equation}
{\bf \mu} = \mu_+ \hat{\bf \varepsilon}_+ 
               + \mu_- \hat{\bf \varepsilon}_- 
               + \mu_z \hat{\bf z},
\end{equation}
and likewise for $\mathbf{B}$ and $\mathbf{U}$ with
$\hat{\bf \varepsilon}_\pm = ( \hat{\mathbf{x}} \pm
\hat{\mathbf{y}})/\sqrt{2}$. The hyperfine transition resonance
frequency $\Omega \equiv \{ H^{\mathrm{atom}}_{22} -
H^{\mathrm{atom}}_{11} - \mu_z^{F=2} B_z^{DC}(\mathbf{r}_j) +
\mu_z^{F=1} B_z^{DC}(\mathbf{r}_j) \} / \hbar$ can be tuned with an
applied DC magnetic field along the quantization axis.
Furthermore, given $\mu_+ = \mu_- \equiv \mu$, it is convenient to
define the Rabi frequency coefficient $\chi_{\lambda}(\mathbf{r}_j) \equiv \mu
\sqrt{\mu_0 \hbar \omega_{\lambda}/2}\,
U_{\lambda}(\mathbf{r}_j)/\hbar$ and employ Pauli matrix operators in
writing the Hamiltonian as:
\begin{eqnarray}
\label{Hamiltonian2}
  H &=& \sum_{\lambda} \hbar \omega_{\lambda} \left( \hat{a}_{\lambda}^{\dag}\hat{a}_{\lambda} + 1/2 \right)+ \\
\sum_{j=1}^N & & \left\{\frac{1}{2} \hbar \left(\Omega - \mathbf{k} \cdot \mathbf{v}_j \right)
  \left( \hat{\sigma}_{0j} + \hat{\sigma}_{zj} \right)-i \hbar \sum_{\lambda}
  \left[  \chi_{\lambda}(\mathbf{r}_j)\hat{\sigma}_{+j}\hat{a}_{\lambda} -
  \chi_{\lambda}^*(\mathbf{r}_j)\hat{\sigma}_{-j}\hat{a}_{\lambda}^{\dag}\right]\right\}.\nonumber
\end{eqnarray}
Here it is understood that the field operators represent only right circular polarization 
and the atom operators act only on operators of the same atom.  
The Heisenberg equations of motion resolve to:
\begin{eqnarray} \label{eq:szdot}
  \dot{\hat{\sigma}}_{zj} &=& -2 \sum_{\lambda} \left\{
  \chi_{\lambda}(\mathbf{r}_j)\hat{\sigma}_{+j}\hat{a}_{\lambda} +
  \chi_{\lambda}^*(\mathbf{r}_j)\hat{\sigma}_{-j}\hat{a}_{\lambda}^{\dag}
  \right\} \\ \label{eq:sminusdot}
  \dot{\hat{\sigma}}_{+j} &=& i \left( \Omega - \mathbf{k} \cdot \mathbf{v}_j \right)\hat{\sigma}_{+j} + \sum_{\lambda}
  \chi_{\lambda}^*(\mathbf{r}_j)\hat{\sigma}_{zj}\hat{a}_{\lambda}^{\dag} \\ \label{eq:adot}
  \dot{\hat{a}}_{\lambda} &=& -i \omega_{\lambda}
  \hat{a}_{\lambda} + \sum_{j=1}^N \chi_{\lambda}^*(\mathbf{r}_j)\hat{\sigma}_{-j}.
\end{eqnarray}
Note that total energy, excitation number are conserved in a lossless system as well as the Bloch vector length.
Optical losses from the cavity can be included by adding a negative imaginary loss rate, 
$\Gamma=\omega_{\lambda}/2Q$, to the mode frequency.  These equations suggest the operators can be
factored as $ \hat{a}_{\lambda} =\tilde{a}_{\lambda} e^{-i \omega_{\lambda} t}$ and
$\hat{\sigma}_{\pm j} = \tilde{\sigma}_{\pm j} e^{\pm i\left(\Omega - \mathbf{k}
\cdot \mathbf{v}_j \right) t}$, isolating the variable envelopes which are easier to integrate numerically.
Furthermore, if the cavity lifetime is much shorter than the superradiant lifetime ($1/\Gamma \ll T_{\mathrm{SR}}$),
then it is convenient to first integrate Eq.~(\ref{eq:adot}) approximately through a Green's function method by
assuming the atom does not evolve significantly within one cavity lifetime \cite{Eberly}:
\begin{equation}\label{eq:aintg}
    \hat{a}_{\lambda}(t) \approx \tilde{a}_{\lambda}(0) e^{-(i \omega_{\lambda}+\Gamma) t} + \sum_{j=1}^N
\chi_{\lambda}^*(\mathbf{r}_j) g_{\lambda j}(t)\,\hat{\sigma}_{-j}, \;\;\;\;\;\; \mathrm{for}\;\;\; \Gamma\, T_{SR} \gg 1,
\end{equation}
where $g_{\lambda j}(t)\equiv (1-\exp[(-i \Delta_{\lambda j} - \Gamma) t])/(i \Delta_{\lambda j} +
\Gamma)$ with detuning between cavity and atom resonance is defined as
$\Delta_{\lambda j} \equiv \omega_{\lambda}-\left(\Omega - \mathbf{k} \cdot \mathbf{v}_j \right)$.
Essentially, each atom responds to the  emission from all atoms accumulated over one cavity lifetime.
This result can then be inserted into Eqs.~(\ref{eq:szdot}) and (\ref{eq:sminusdot}),
defining the detuning between atom resonances $\Delta_{i j} \equiv \mathbf{k} \cdot (\mathbf{v}_i-\mathbf{v}_j)$, to yield:
\begin{eqnarray} 
  \dot{\tilde{\sigma}}_{zj} &=& -2 \sum_{\lambda} \Bigg\{
  \chi_{\lambda}(\mathbf{r}_j) e^{(-i \Delta_{\lambda j} - \Gamma) t} \tilde{\sigma}_{+j}\tilde{a}_{\lambda}(0) +
  \chi_{\lambda}^*(\mathbf{r}_j) e^{(i \Delta_{\lambda j} - \Gamma) t} \,
\tilde{\sigma}_{-j}\tilde{a}_{\lambda}^{\dag}(0) \\
  & + & \sum_{i=1}^N   \left\{
  \chi_{\lambda}(\mathbf{r}_j) \chi_{\lambda}^*(\mathbf{r}_i)g_{\lambda i}(t) e^{-i \Delta_{i j} t}
\tilde{\sigma}_{+j}\tilde{\sigma}_{-i} +
  \chi_{\lambda}^*(\mathbf{r}_j)\chi_{\lambda}(\mathbf{r}_i)g^*_{\lambda i}(t) e^{i \Delta_{i j} t}
\, \tilde{\sigma}_{-j}\tilde{\sigma}_{+i}
\right\}\Bigg\}, \nonumber
\label{eq:approxeqnofmotion1} 
\end{eqnarray}
\begin{equation}
\dot{\tilde{\sigma}}_{+j}  =  \sum_{\lambda} \left\{
\chi_{\lambda}^*(\mathbf{r}_j)  e^{(-i \Delta_{\lambda j} - \Gamma) t} \tilde{\sigma}_{zj}
\tilde{a}_{\lambda}(0) + \sum_{i=1}^N \chi_{\lambda}^*(\mathbf{r}_j)\chi_{\lambda}(\mathbf{r}_i)
g^*_{\lambda i}(t)\,e^{i \Delta_{i j} t} \,  \tilde{\sigma}_{zj}\tilde{\sigma}_{-i} \right\}.
\label{eq:approxeqnofmotion2}
\end{equation}
The field intensity, square of Eq.~(\ref{eq:aintg}), can then be produced from the atomic state by first
numerically integrating these reduced equations.  The terms linear in $\tilde{a}_{\lambda}(0)$ represent
the negligible action by the initial field.  The  quadratic terms involve feedback from all other atoms
that generates superradiance.  Note that on resonance ($\Delta_{\lambda j}=0$) we arrive at the expected
time scale, for $t \ll 1/\Gamma$, through the coefficient:
\begin{equation} \label{eq:cavitylifetime}
\chi_{\lambda}^*(\mathbf{r}_j)\chi_{\lambda}(\mathbf{r}_i)g_{\lambda j}(t)= \frac{\mu_0 \,
\mu^2 \, \omega_{\lambda}}
{2 \, \hbar \, V_{\lambda}}\frac{2 Q}{\omega_{\lambda}} V_{\lambda}\,
U_{\lambda}^*(\mathbf{r}_j)\, U_{\lambda}(\mathbf{r}_i) = \frac{1}{2 \, T_1^\mathrm{cav}}
 V_{\lambda}\, U_{\lambda}^*(\mathbf{r}_j)\, U_{\lambda}(\mathbf{r}_i).
\end{equation}

For significant amplification, the number of interacting atoms must be large and 
it becomes prohibitively difficult to solve the $2N$ coupled equations of motion.  
Instead one can dice the atomic distribution in phase space into finite close-packed 
cells so that the number of cells is computationally feasible and the pertinent
distribution and field profile features are retained. These constraints typically 
imply that the number of atoms per cell is at least ten. 
In this case, the correlation between the atomic population and photon number 
is inconsequential (i.e. atom stimulation rate is not significantly limited 
by the stored energy) and the expectation value of the operator products can be factored as 
$<\tilde{\sigma}_{+j}\tilde{a}_{\lambda}> \approx <\tilde{\sigma}_{+j}><\tilde{a}_{\lambda}>$.  
The sum over atoms can then be replaced with the multiple sum over cell indices:
\begin{equation}
\sum_{j=1}^N f_j(\mathbf{r}_j,\mathbf{v}_j) \rightarrow \sum_i \sum_j \sum_k \sum_l
N_{ijkl}\, f_{ijkl}(<x>_i^{\mathrm{cell}},<y>_j^{\mathrm{cell}},<z>_k^{\mathrm{cell}},<v_{\|}>_l^{\mathrm{cell}}),
\end{equation}
provided the atom position and velocity are replaced with the mean cell position 
$\mathbf{r}_j \rightarrow <\mathbf{r}>_{cell}$ and velocity 
$\mathbf{v}_j \rightarrow <\mathbf{v}>_{cell}$ and the atom operators 
now represent the typical atom in the cell. 
The number of atoms in cell $\{i,j,k,l\}$ is designated as $N_{ijkl}$, and 
only the parallel component of velocity is relevant. The cellular equations 
of motion can then be numerically integrated given the initial expectation 
values, atom and field distribution, atom resonance and moment, and field resonance.
Collisions can be simulated by introducing a random phase factor, depending on 
the number of atoms in the cell, to the atom operators between time steps.
Assuming an initially inverted atomic population, superfluorescence can be simulated by assuming zero
photons present and an initial "tipping angle", $<\tilde{\sigma}_{zj}>(t=0)=2/\sqrt{N_{\mathrm{at}}}$,
of the Bloch vector representing spontaneous atom state fluctuations \cite{Haroche}.
Likewise, Casimir stimulated superradiance may be simulated by assuming the same tipping angle
and initial field amplitude $<\tilde{a}_{\lambda}>(t=0)=\sqrt{N_{\mathrm{Cas}}^{\mathrm{max}}}$,
given the parametric amplification process generates a coherent state of the field.
\begin{figure}[t]
\begin{center}
\includegraphics[width=0.9\columnwidth,clip]{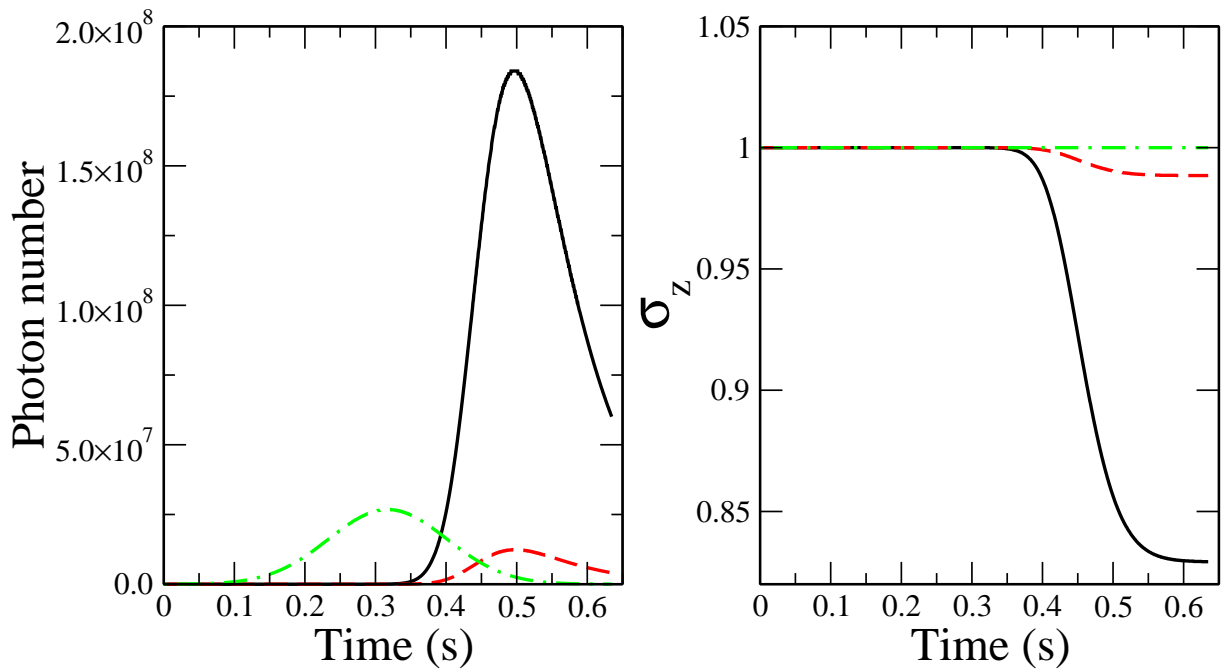}
\caption{Dynamics of superradiant amplification of Casimir photons (assumed in the number of $10^2$) 
and competition with superfluorescence. On the left plot, the superradiant emitted photons  
(solid line, black) compared to the superfluorescence in the absence of Casimir photons (dotted line, red) 
are depicted versus time. The dot-dashed curve in green indicates the field strength profile felt by the atoms.
On the right plot, the atomic inversion populations for superradiant (solid line, black) and 
superfluorescence (dotted line, red) are also plotted versus time. The dot-dashed curve 
curve in green indicates the degree of conservation of the Bloch vector length during the simulation. 
The simulation is performed for for Na atoms with $Q=10^9, N_{at}=10^{10}$, 1 m/s atom velocity, 
10 mK temperature. The relevant time scales are: $1/\Gamma=0.18$s, $T_1^{\mathrm{cav}}=10^6$s, 
$T_{\mathrm{SR}}=10^{-4}$s, and Doppler dephasing time $T_2^*=63$ ms.}
\end{center}
\label{fig2}
\end{figure}

\begin{table}[htbp] 
\begin{center}
\centering
\begin{tabular}{ @{} ccccccccccccc @{} }
\hline
$N^{\mathrm{cas}}_{\mathrm{max}}$ & $N_{\mathrm{at}}$ & $v_{\mathrm{at}}$ (m/s) & $Q$ 
& $T_{\mathrm{at}}$ (K) & $\left<a^{\dagger}a\right>$ (peak) 
& $\eta$ & $\rho_{\mathrm{gnd}}^{\mathrm{SR}}$ (\%) & $\xi$ \\
\hline
\hline
$10^2$ & $1 \times 10^{10}$ & 1  & $1 \times 10^9$ & 0.01 & $1.8 \times 10^8$    & 15.  & 8.5                  & 15  \\
$10^2$ & $1 \times 10^{10}$ & 1  & $2 \times 10^9$ & 0.01 & $9.6 \times 10^8$    & 26.  & 34.                  & 26  \\
$10^2$ & $1 \times 10^{10}$ & 1  & $4 \times 10^9$ & 0.01 & $2.1 \times 10^9$    & 31.  & 39.                  & 20 \\
$10^2$ & $5 \times 10^9   $ & 1  & $1 \times 10^9$ & 0.01 & $5.1 \times 10^5$    & 14.  & 0.049                & 14. \\
$10^2$ & $2 \times 10^{10}$ & 1  & $1 \times 10^9$ & 0.01 & $8.2 \times 10^9$    & 1.2  & 49.                  & 1.8 \\
$10^2$ & $4 \times 10^{10}$ & 1  & $1 \times 10^9$ & 0.01 & $2.4 \times 10^{10}$ & 1.0  & 23.                  & 7.3 \\
$10$   & $1 \times 10^{10}$ & 1  & $1 \times 10^9$ & 0.01 & $4.5 \times 10^7$    & 3.6  & 2.1                  & 3.6 \\
$1$    & $1 \times 10^{10}$ & 1  & $1 \times 10^9$ & 0.01 & $2.0 \times 10^7$    & 1.6  & 0.95                 & 1.6 \\
$10^2$ & $1 \times 10^{10}$ & 1  & $1 \times 10^9$ & 0.1  & $1.0 \times 10^8$    & 15.  & 4.6                  & 15. \\
$10^2$ & $1 \times 10^{10}$ & 1  & $1 \times 10^9$ & 1    & $9.4 \times 10^7$    & 15.  & 4.3                  & 15. \\
$10^2$ & $1 \times 10^{10}$ & 1  & $1 \times 10^9$ & 10   & $9.3 \times 10^7$    & 15.  & 4.3                  & 15. \\
$10^2$ & $2 \times 10^{10}$ & 2  & $1 \times 10^9$ & 0.01 & $6.3 \times 10^6$    & 30.  & 0.11                 & 30. \\
$10^2$ & $2 \times 10^{10}$ & 2  & $2 \times 10^9$ & 0.01 & $1.8 \times 10^7$    & 45.  & 0.26                 & 45. \\
$10^2$ & $2 \times 10^{10}$ & 2  & $4 \times 10^9$ & 0.01 & $3.0 \times 10^7$    & 56.  & 0.4                  & 56. \\
$10^2$ & $1 \times 10^{10}$ & 2  & $1 \times 10^9$ & 0.01 & $6.7 \times 10^4$    & 28.  & $2.5 \times 10^{-3}$ & 28. \\
$10^2$ & $4 \times 10^{10}$ & 2  & $1 \times 10^9$ & 0.01 & $5.4 \times 10^9$    & 22.  & 48.                  & 22. \\
$10^2$ & $1 \times 10^{11}$ & 10 & $1 \times 10^9$ & 0.01 & $4.2 \times 10^4$    & 760. & $7 \times 10^{-5}$   & 680 \\
$10^2$ & $4 \times 10^{11}$ & 10 & $1 \times 10^9$ & 0.01 & $2.9 \times 10^8$    & 650. & 0.11                 & 650 \\
$10^2$ & $1 \times 10^{12}$ & 10 & $1 \times 10^9$ & 0.01 & $8.7 \times 10^{11}$ & 42.  & 25.                  & 8.6 \\
$10^2$ & $1 \times 10^{11}$ & 10 & $2 \times 10^9$ & 0.01 & $5.3 \times 10^4$    & 840. & $8.3 \times 10^{-5}$ & 760 \\
$10^2$ & $1 \times 10^{11}$ & 10 & $4 \times 10^9$ & 0.01 & $6.0 \times 10^4$    & 890. & $9.1 \times 10^{-5}$ & 810 \\
\hline
\end{tabular}
\caption{Configurations yielding effective discrimination between Casimir superradiant signals
and superfluorescence noise. The degree of discrimination is indicated by the ratio of the peak 
intensity for the SR case to the SF case ($\eta$) and the analogous ratio of the ground state population 
after the atoms exit the cavity ($\xi$).  The ground state population of the atoms near resonance is 
given by $\rho_{\mathrm{gnd}}^{\mathrm{SR}}=(1-\left<\sigma_z(\Delta\approx 0)\right>)/2$. 
The superradiant lifetime for these cases is in the range of $T_{\mathrm{SR}}=1-100 \mu$s, 
roughly $10^3$ times less than the cavity decay time, implying $\Gamma T_{\mathrm{SR}} \ll 1$. }
\end{center}
\end{table}

\section{Discussion of numerical results and conclusions}

The coupled equations of motion, Eqs.~(\ref{eq:szdot})-(\ref{eq:adot}), were integrated by
Runge-Kutta method for a small sample of Na atoms traversing a planar-concave optical
cavity through the diameter at 10$\%$ of the cavity length height above the planar mirror.
The cavity is tuned so that the first longitudinal TEM00 Gaussian mode is resonant with the atoms.
The atom distribution was divided into five cells varying in velocity symmetrically about the mean speed.
The number of atoms, cavity quality, atom mean speed, atom transverse temperature (i.e. velocity spread),
and initial photon population were varied to identify promising configurations and sensitivity of the 
superradiant output. We found that there is a range of configurations, corresponding to the high 
end of experimental feasibility, that achieve a high degree of discrimination between the Casimir 
stimulated superradiant and the superfluorescent emission. In Fig. 2  we show the typical emission profile 
(left panel) and atomic inversion population (right panel). As suggested above, the superfluorescent 
emission is suppressed if the atoms leave the field mode region before its amplification can develop. 
Even a single photon in the cavity can generate a signal 60 $\%$ above the superfluorescent background. 
Atom numbers too high or too low, depending primarily on $Q$ and atom speed, allow both cases to 
develop fully or very little, respectively. Superradiant amplification is quite effective for 
sufficiently high gain, easing RF detector sensitivity requirements. The gain is very sensitive 
to $N_{at}$ and $Q$ and the atom transit time ($\propto 1/v_{\mathrm{at}}$). 
Also, the strong discrimination in amplification indicated by $\eta$ in Table 1 implies that the 
detector response time need not be a significant constraint, as would be the case if the delay time 
should be measured.  The discrimination ratio is fairly insensitive to temperature, suggesting 
that a slowed beam rather than a trap could be used, providing larger $N_{at}$.  
Strong amplification necessarily results in a large fraction of the atoms driven to their ground state. 
Therefore, probing the atomic beam ground state through D-line excitation after it transits the cavity, 
as we had proposed in \cite{Andy}, is a viable option that avoids detector losses in the cavity.
The high values of $N_{at}$ and $Q$ would be eased significantly if the number of initial Casimir photons 
is increased.  Equation~(\ref{eq:sat}) implies $N_{\mathrm{max}}^{\mathrm{Cas}}={0.7,6,370,10^6}$ for 
$Q \epsilon = {0.5,1,2,3,4}$.  The rapid change in initial photon population $N_{\mathrm{max}}^{\mathrm{Cas}}$ 
due to $Q$ for a given modulation mechanism is not taken into account in Table 1. 
One caveat to the above results and discussion is that the computed time for the superradiant 
burst to develop is far longer than the standard prediction, Eq.~(\ref{eq:delay}). 
The configurations tested imply $\Gamma\, T_{SR} \ll 1$, in violation of the approximation condition under which
Eq.~(\ref{eq:delay}) was derived. Indeed, the reduced equations of motion, Eqs.~(\ref{eq:aintg})
and (\ref{eq:approxeqnofmotion2}), were quite unstable under these conditions. As we have shown, the standard 
result develops naturally from the general theory given a lossy environment. The transition to a low 
loss environment appears to have a dramatic effect delaying the field growth. We are currently studying 
the origin and domain of validity of this effect.

\vspace{0.5cm}
RO acknowledges partial support by the NSF through the Institute for Theoretical Atomic and Molecular 
Physics at Harvard University and the Smithsonian Astrophysical Observatory.

\end{document}